\documentclass[aps,prl,preprintnumbers,amsmath,amssymb,twocolumn,superscriptaddress,showpacs,floatfix]{revtex4}
\usepackage{epsfig}
\usepackage{dcolumn}
\usepackage{bm}

\newcommand{\ave}[1]{\left\langle #1 \right\rangle}

\begin{document}
\title{Polarization probes of vorticity in heavy ion collisions}


\author{Barbara Betz}
\affiliation{Institut f\"ur Theoretische Physik, J.W. Goethe-Universit\"at, Frankfurt, Germany}
\affiliation{Helmholtz Research School, Universit\"at Frankfurt, GSI and FIAS, Germany}
\author{Miklos Gyulassy}
\affiliation{Institut f\"ur Theoretische Physik, J.W. Goethe-Universit\"at, Frankfurt, Germany}
\affiliation{Frankfurt Institute for Advanced Studies (FIAS), Frankfurt, Germany}
\affiliation{Department of Physics, Columbia University, New York, New 
York, 10027, USA}
\author{Giorgio Torrieri}
\affiliation{Institut f\"ur Theoretische Physik, J.W. Goethe-Universit\"at, Frankfurt, Germany}
\affiliation{Frankfurt Institute for Advanced Studies (FIAS), Frankfurt, Germany}
\date{June 2007}

\begin{abstract}
We discuss the information that can be deduced from a measurement of
hadron (hyperon or vector meson) polarization in ultrarelativistic nuclear collisions.
We describe the sensitivity of polarization to initial conditions, hydrodynamic evolution and mean 
free path, and find that the polarization observable is sensitive to all details and stages of the system's evolution.   We suggest that an experimental investigation
covering production plane and reaction plane polarizations, as well as the polarization of jet-associated particles in the plane defined by the jet and particle direction,  can help 
in disentangling the factors contributing 
to this observable.   Scans of polarization in energy and rapidity might also 
point to a change in the system's properties.
\end{abstract}


\maketitle

Parity violation, together with the self-analyzing nature of hyperon decays, provides us with the 
opportunity to study the polarization of hyperons produced in heavy ion collisions:  In the rest frame 
of the hyperon $Y$, the angular decay distribution w.r.t. the polarization plane is \cite{bunce}
\begin{equation}
\label{poldef}
\frac{dN}{d \theta} = 1 + \alpha_Y P_Y \cos(\theta)
\end{equation}
where $\alpha_Y$ is a hyperon-specific constant (measured in elementary processes \cite{bunce}), $P_Y$ is the hyperon polarization and $\theta$ is the angle between the proton momentum and $\Lambda$ polar axis.     Similar relations arise, without the need for parity violation, in the strong decays of vector mesons to ``p-wave'' final states \cite{jacob}.

If a specific net $\ave{P_Y} \ne 0$ exists in any axis definable event-by-event, it is in principle possible to measure it 
using Eq. \ref{poldef} and the observed spectra of $\Lambda,\Xi,\Omega$ decay products.
This opens a new avenue to investigate heavy ion collisions, which has been proposed both as a signal of a 
deconfined regime \cite{hoyer,stock,Panagiotou:1986zq,rafpol} and as a mark of global properties of the event \cite{liang,liangvector,becattini,qm2006}.

Since QCD contains a spin-orbit coupling, a non-zero hyperon polarization in direction $i$ is in principle present whenever the angular momentum density in that direction, 
$\ave{\vec{x} \times \vec{ T^{0 i}}}_i$, is non-zero ($T^{\mu \nu}$ is here the energy momentum tensor).  
As we will see, in both elementary p-p, p-A and A-A collisions is is possible to define directions where this vector might have non-zero components.

\begin{figure*}[t]
\epsfig{width=18cm,clip=,figure=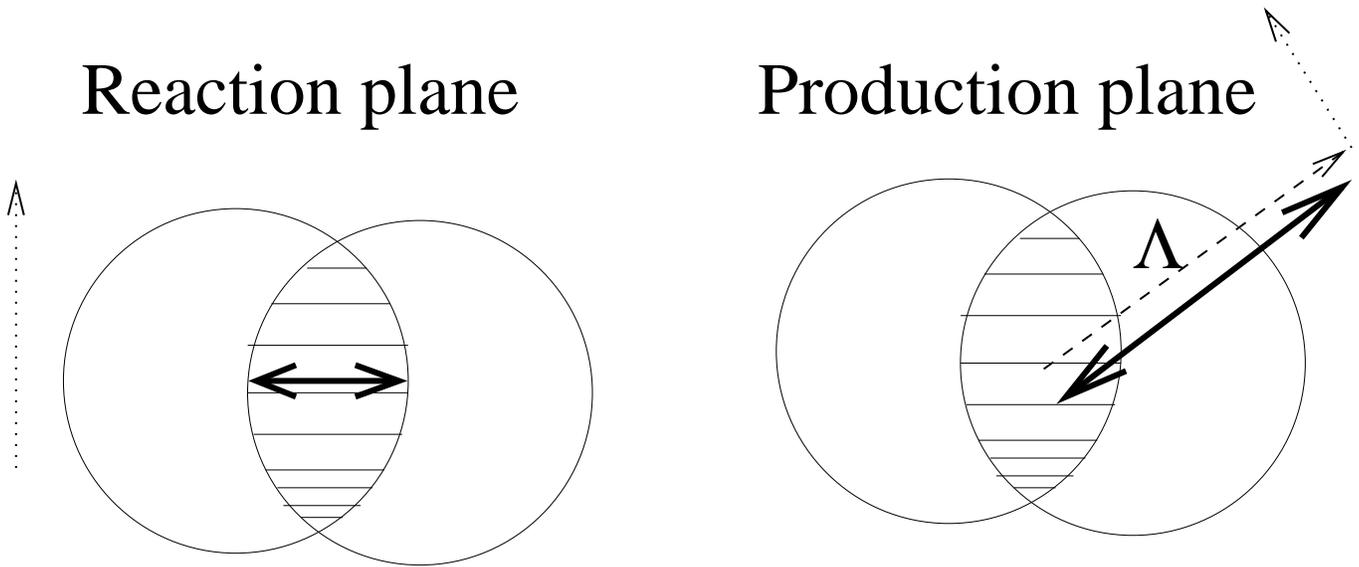}
\caption{\label{planes} (color online) Definition of production and reaction plane.  The beam line (traditionally the $z$ axis) 
is perpendicular to the sheet.  The dotted line, with arrow, indicates 
the direction of polarization of the produced $\Lambda$. }
\end{figure*}

The potential of hyperon polarization as a signal for deconfinement comes from the  strong transverse polarization 
of hyperons in the production plane (left panel of fig. \ref{planes}) observed
in (unpolarized) $p+p$ and $p+A$ collisions \cite{PhData,boros}.
As suggested in \cite{hoyer,stock,Panagiotou:1986zq}, the disappearance of this polarization (which we shall call $P_{Y}^P$) could signal the 
onset of an isotropized system where, locally, no reference frame is preferred.  Something close to what is today 
called a ``Quark Gluon liquid'' ({\underline s}trongly interacting {\underline Q}uark-{\underline G}luon 
{\underline P}lasma, sQGP).    So far, no such measurement exists at RHIC energies, through an Alternating Gradient Synchrotron (AGS) measurement \cite{agsprod} yields a negative result (transverse polarization is comparable to p-A collisions).

It has also been suggested \cite{liang} to use hyperon polarization in the
{\em reaction plane} (left panel, Fig. \ref{planes}) to test for local angular momentum in the matter produced in 
heavy ion collisions.
The idea is that the initial momentum gradient in non-central collisions should result in a net angular momentum (shear) in this direction, that will be transferred to hyperon spin via spin-orbit coupling (This polarization direction will be called $P_{Y}^R$).   A similar, through quantitatively different result, can be obtained from a microcanonical ensemble with a net angular momentum \cite{becattini}.

The STAR collaboration has recently measured the reaction polarization \cite{starpol}, reporting results consistent 
with zero.   The production plane polarization measurement is also planned.

In this paper, we make a few general considerations regarding the insights that can be gained from polarization measurements.   We examine how the 
polarization, in both production and reaction plane, is sensitive to initial conditions, hydrodynamic evolution, 
and mean free path.
We suggest measuring polarization in different directions (production, reaction, and jet axis in jetty events) 
could provide a way to go beyond model-dependence.

While throughout this paper we use the $\Lambda$ polarization as our 
signature of choice, the points made here can be easily generalized to 
the detection of polarization of vector mesons, also used as probes of polarization in a very similar way to hyperons \cite{liangvector}

\vspace*{-0.2cm}
\section{Initial conditions and reaction plane polarization}
The QCD spin orbit coupling is capable of transforming the orbital angular momentum {\em density} $\ave{\vec{x} \times \vec{T^{0 i}}}_i = \ave{\vec{x} \times \vec{p}}$ into spin.  
The fact that the relevant quantity is angular momentum {\em density}, rather than its absolute value, can be seen intuitively by the requirement of locality.   Formally, it is apparent if the polarization from scattering is calculated explicitly using a Wigner Function formalism \cite{liang}.   
For a large system, such as a heavy nucleus, we have to convolute the net polarizing interaction cross-section per unit of transverse nuclear surface
 ($ d \Delta \sigma/d^2 x_\perp$, where $x_\perp$ are the two directions perpendicular to the beam axis) calculated in \cite{liang} with the (initial) parton phase space distributions $f(x_\perp,p)$ to obtain the net local polarized parton phase space density $\rho_{P^R_q}$ produced in the first interactions
\begin{equation}
\rho_{P^R_q} (x_\perp,p) =  \int d^2 x'_\perp d^3 p' f(x_\perp-x_\perp',\vec{p}-\vec{p}') \frac{ d \Delta \sigma}{d^2 x_\perp'} \left(p' \right) 
\end{equation}
where $f(x_\perp,p)$ is the local parton distribution of the medium.

Provided the initial Debye mass and constituent quark mass is small,  the quark polarization in the reaction plane $$\ave{P_q^R} = \int d^2 x_\perp d^3 p \rho_{P^R_q} (x_\perp,p)$$ becomes \cite{liang}


\begin{equation}
\ave{P_q^R} \sim \int d^2 x_\perp \rho(x_\perp) \vec{p}.(\vec{x_\perp} \times \vec{n}) \sim - \ave{ p_z x_\perp}   
\label{pqint}
\end{equation}
where  $$\rho(x_\perp)  =\int d^3 p f(x_\perp,p) $$ is the participant transverse density and $\vec{n}$ is a unit vector perpendicular to both $x_\perp$ and $\vec{p}$.   In ultrarelativistic collisions
all significant initial momentum is in the beam ($z$) direction.

In non-central collisions with a finite impact parameter $\vec{b}$, $\ave{\vec{p} \times \vec{x_\perp}} \propto \vec{b} \ne 0$, thereby generating a resulting net polarization. 
 
Thus, the initially generated amount of reaction plane polarization is strongly dependent on the initial density-momentum correlation within the system.  In other words, the reaction plane polarization could be a useful signature for probing the initial conditions within the system created in heavy ion collisions.

According to the Glauber model, the initial density transverse coordinate distribution is given by the sum of the participant and target density $\rho_{P,T}$.
\begin{equation}
\rho(x_\perp) = \left( \rho_{P} (x_\perp) + \rho_T (x_\perp) \right) \phi (y,\eta)
\end{equation}
where
\begin{equation}
\rho_{p,T} = T_{p,T} \left( x_\perp \mp \frac{b}{2} \right) \left( 1 - \exp \left[ -\sigma_N T_{T,p} \left( x_\perp \pm \frac{b}{2} \right) \right]  \right)
\end{equation}
and $\sigma_N$, $T_{P,T}$ and $b$ refer, respectively, to the nucleon-nucleon cross-section, the nuclear (projectile and target) density, and the impact parameter.

\begin{figure*}[t]
\epsfig{height=18cm,clip=,figure=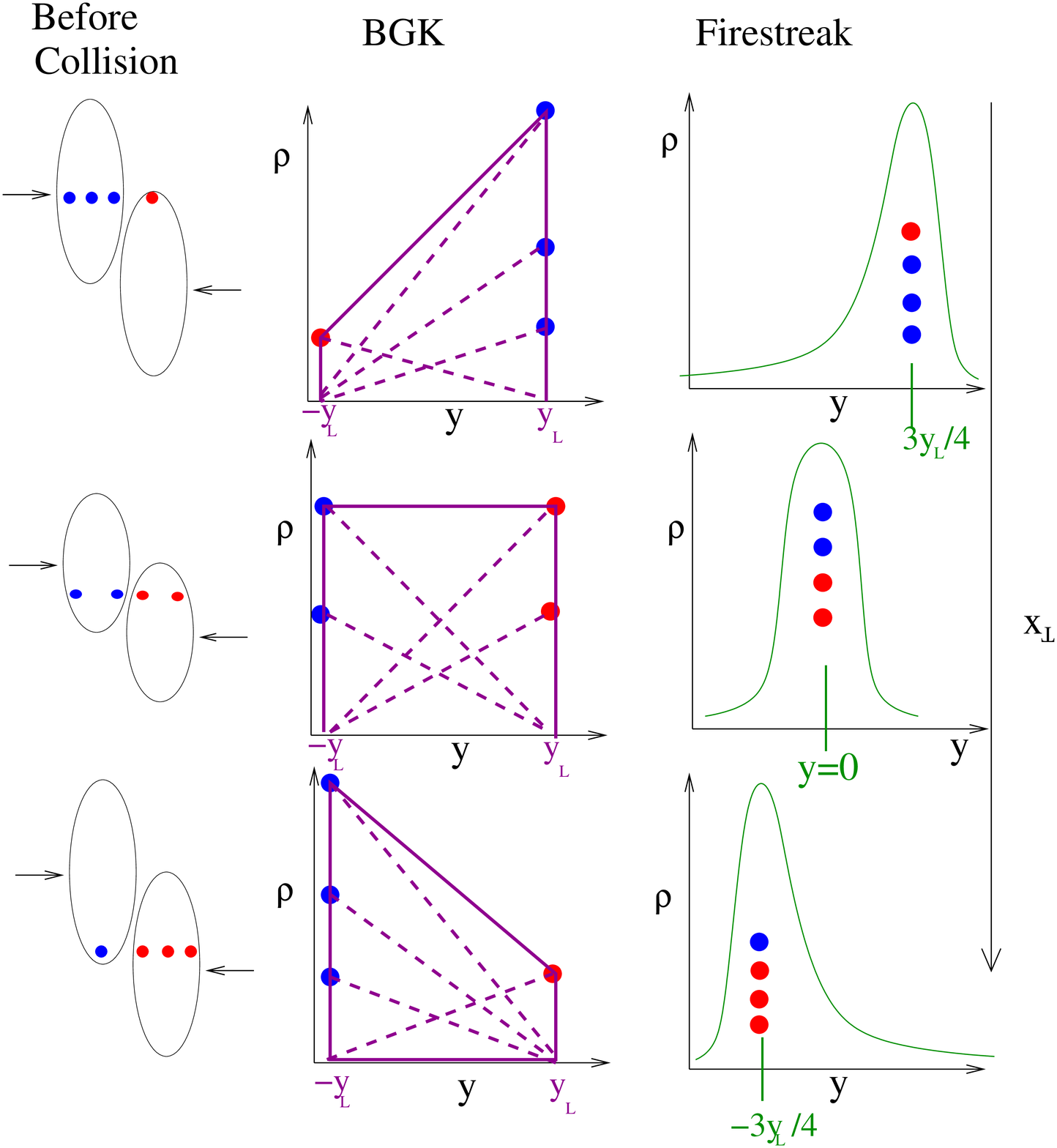}
\caption{\label{initial_density} (color online) Initial densities in the BGK model (left), as well as
the model used in \cite{liang} (right).
In the BGK case, dashed lines represent the rapidity extent of the ``excited state'' produced by the individual nucleon, while solid lines correspond to the cumulative density.  See text for model definitions and further explanations}
\end{figure*}

How this density is longitudinally distributed in spacetime rapidity
\begin{equation}
\label{spacerap}
\eta = \frac{1}{2} \ln \left( \frac{t+z}{t-z}  \right)
\end{equation}
and flow rapidity
\begin{equation}
\label{flowrap}
y =  \frac{1}{2} \ln \left( \frac{E+p_z}{E-p_z}  \right) = \frac{1}{2} \ln \left( \frac{1+v_z}{1-v_z}  \right)
\end{equation}
(the form of $\phi (y,\eta)$) is a crucially important model parameter.
The calculation of the hyperon polarization in the reaction plane \cite{liang} is dependent on an
assumption of initial condition described in Fig. 1 and Eq. 1 of
\cite{liang}. Such an initial condition (generally referred to as the ``firestreak model''), is roughly equivalent 
to two
``pancakes'', inhomogeneous in the transverse coordinate $x_\perp$, locally
inelastically sticking together.   Each element of this system then
streams in the direction of the local net momentum (Fig. \ref{initial_density}, right column).
   
Since projectile and target have opposite momenta in the
center of mass frame, assuming projectile and target nuclei to be identical yields
\begin{equation}
\phi (y,\eta) \simeq \delta(\eta) \delta \left( y - y_{cm} (x_\perp) \right)
\end{equation}
where $y_{cm}$ is the local (in transverse space) longitudinal rapidity, corresponding to the flow velocity $v_{cm}$ given by momentum conservation.  Thus
\begin{equation}
\label{polawang}
\ave{ p_z x_\perp} \sim \frac{\sqrt{s}}{m_N} \ave{D_\rho}
\end{equation}
where
\begin{equation}
\label{drho}
\ave{D_\rho}= \int d^2 x_\perp x_\perp \left[ \rho_P (x_\perp) - \rho_T (x_\perp) \right] 
\end{equation}
If the colliding nucleons are also the interacting degrees of freedom, then the constant would be independent of energy.    However, we know that at high energy the physical degrees of freedom are partons, and the amount of partons each nucleus ``fragments'' into is highly energy dependent.    \cite{liang} takes this into account using the energy-dependent parameter $c(s)$.
 If one assumes all entropy to be created in the initial moment, $c(s)$ as a function of energy can be estimated from final multiplicity using the well-known phenomenological formulae \cite{milov} ($\sim 18$ at top RHIC energies).
\begin{equation}
\label{csmilov}
c(s) \sim \frac{2 y_{L}}{N_p} \frac{dN}{dy} 
  \simeq \frac{1}{1.5}  \ln \left( \frac{\sqrt{s}}{1.5 GeV} \right)\ln \left( \frac{2 \sqrt{s}}{ GeV} \right) 
\end{equation}
Assuming that all partons receive an equal share of momentum, we get
\begin{equation} 
 \ave{ p_z x_\perp} \sim  \ave{D_\rho} \frac{\sqrt{s}}{c(s) m_N}  
\label{csdef}
\end{equation}
Since all nuclei have the same $\sqrt{s}$, $\ave{P_q^R}$ should be finite and constant over rapidity (Fig. 1 of \cite{liang}).

 The physical validity of such a picture is compelling at
low energies, when the baryon stopping of nuclear matter is large.
At high energies and initial transparencies, however, a more generally accepted ansatz for initial condition is that 
approximated by a Brodsky-Gunion-Kuhn (BGK) \cite{bgk,gyuladil} picture,
where the initial partons are produced all-throughout the 
longitudinal flow rapidity spanned between the forward-traveling projectile and the
backward-traveling target (middle panel of Fig. \ref{initial_density}).     The spacetime rapidity

is, in this picture, equal to the flow rapidity (Hubble/Bjorken expansion).

If $\rho_P (x_\perp)=\rho_T
(x_\perp)$, this reduces to a boost-invariant initial condition.
For a non-central collision, however, such equality will only hold at
the midpoint in $x_\perp$ of the collision region. 
Interpolating linearly in the rapidity $y$ between $\rho_P$ (at $y=y_L$) and $\rho_T$ (at $y=-y_L$), we have
\begin{equation}
\label{rhoybgk}
\phi (y,\eta) =  \left( A  +  y B \right) \delta(y-\eta)
\end{equation}
\begin{equation}
A= \frac{1 }{2},B=\frac{\rho_P (x_\perp) - \rho_T (x_\perp)}{\rho_P (x_\perp) + \rho_T x_\perp}\frac{1}{2 y_L},
\end{equation}
In particular, it means that for identical nuclei
\begin{eqnarray}
\label{pzbgk}
 \ave{p_z x_\perp}  \propto \int d y \sinh(y) x_\perp \rho(x_\perp,y) d x_\perp 
 \nonumber \\ 
\propto \ave{D_\rho} \left( y_L \cosh(y_L) - \sinh(y_L) \right)
\end{eqnarray}
(Note that no $c(s)$ is necessary here since we are dealing with collective longitudinal flow).
For this initial condition, the axial symmetry of the initial pancakes forces the net
polarization at mid-rapidity to be zero, and to rise as $O(y_L^3)$ for $y_L<1$.

\begin{figure*}[t]
\epsfig{width=13cm,clip=,figure=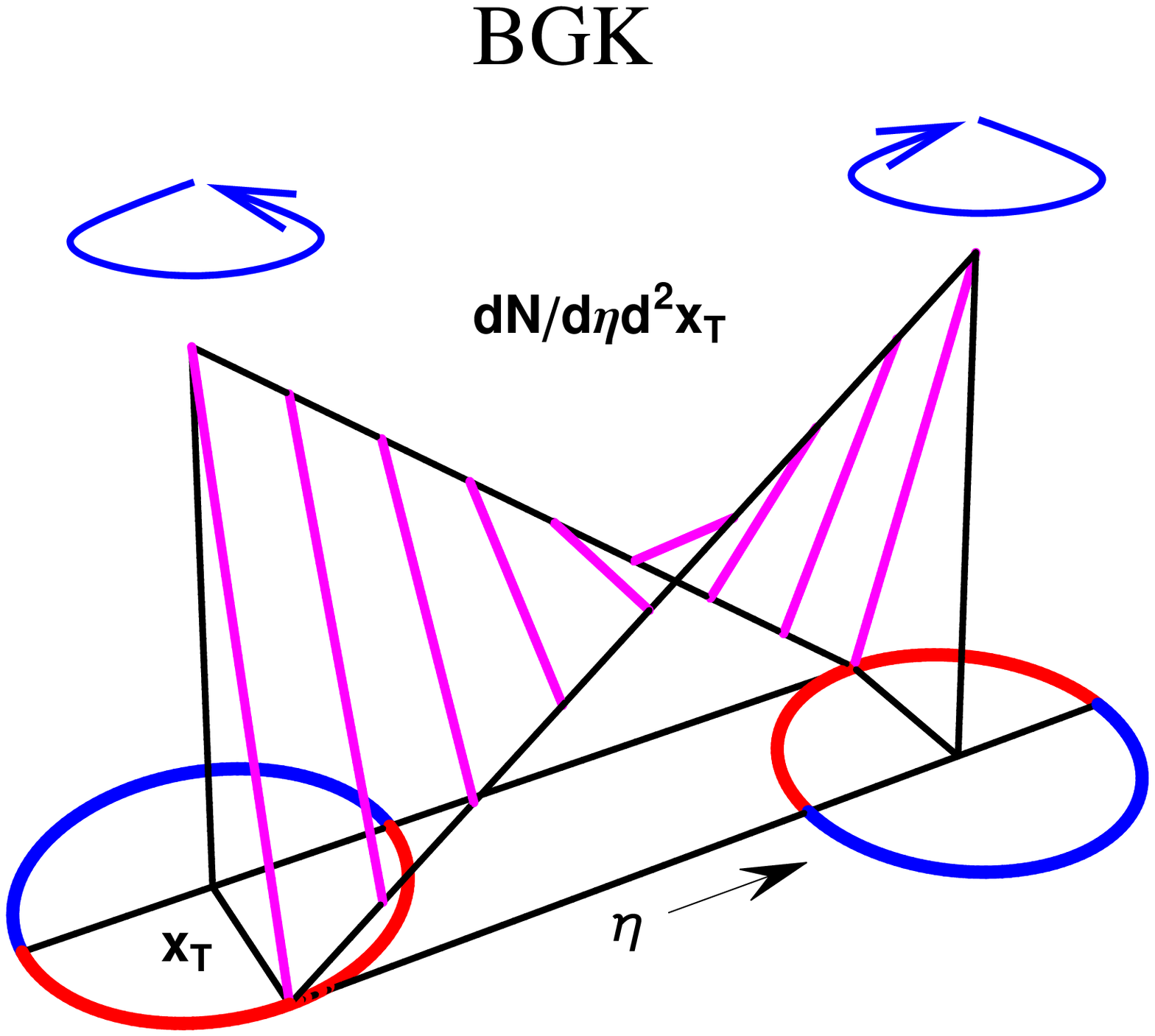}
\epsfig{width=13cm,clip=,figure=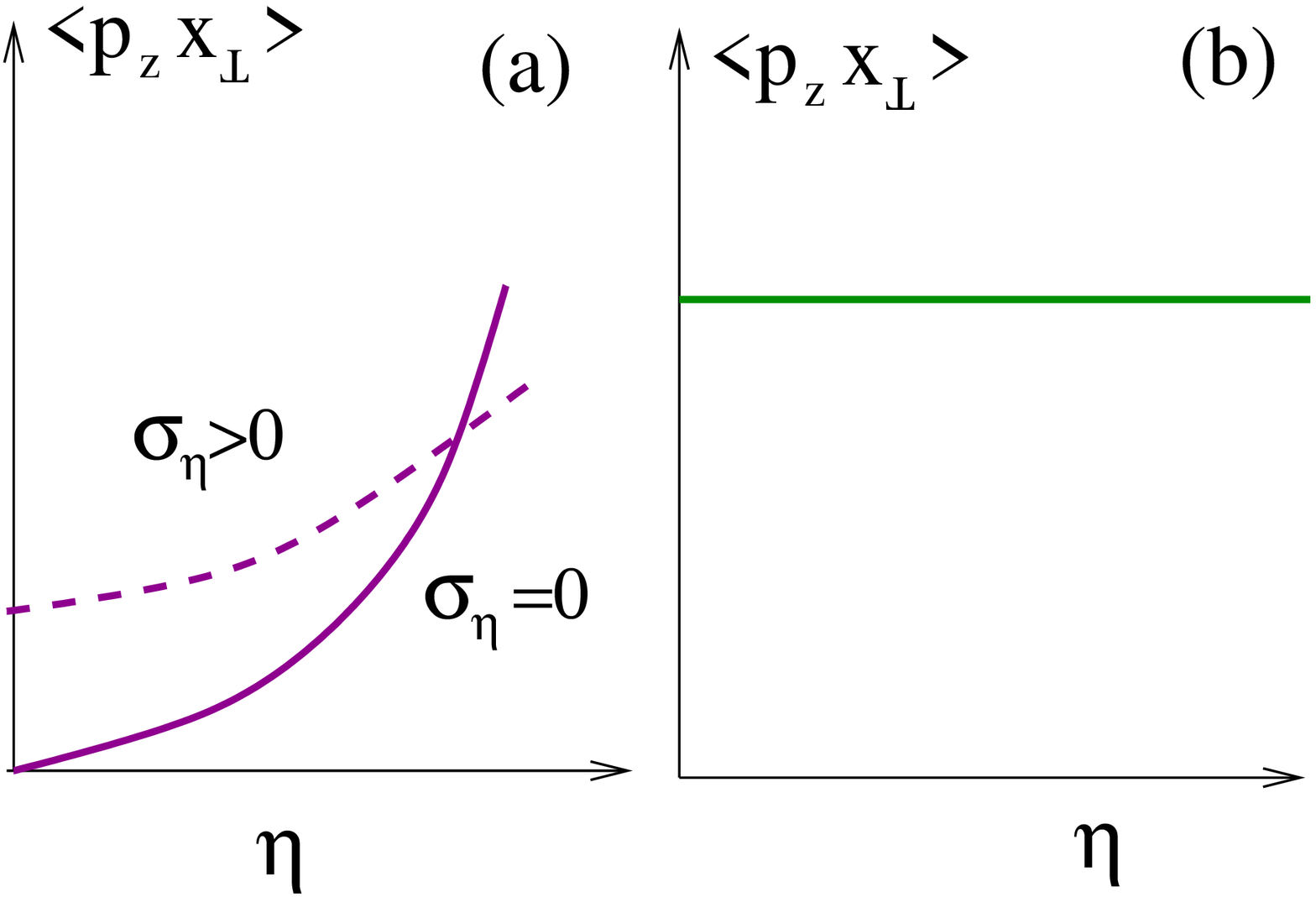}
\caption{\label{initial_shear} (color online) Initial shear in the BK model (a), as well as
the model used in \cite{liang} (b)}
\end{figure*}

The rapidity distributions are
summarized in Fig. \ref{initial_density}, and the corresponding shear
created is summarized in (a) and (b) of Fig. \ref{initial_shear}.

Thus, at high (RHIC and LHC) energies we expect net polarization around
the reaction plane of A-A collisions should vanish at mid-rapidity and
reappear in the target and projectile regions (Panel (a) of Fig. \ref{initial_shear}).
At lower energies, on the
other hand, reaction plane $\Lambda$ or $\overline{\Lambda}$ polarization
should be more uniform in rapidity space, and be significantly above zero at mid-rapidity (Panel (b) of Fig. \ref{initial_shear}).

Realistic nuclear geometries 
should not alter these very basic considerations, although for the BGK
case they might considerably slow down the shear rise in rapidity.
This is also true for corrections to linear interpolation in rapidity space.  
Detailed hydrodynamic simulations \cite{romatshke} also reinforce the conclusion that within the boost invariant limit vorticity is negligible.

A net vorticity will reappear in the case of an ``imperfect'' BGK initial condition where {\em spacetime} rapidity 
and the longitudinal flow rapidity 
are not perfectly correlated. 
Within the ``ideal'' Bjorken limit the correlation is indeed perfect, but its 
reasonable to expect that deviations occur.  

The existence of the ``ridge'' \cite{refridge} provides an experimental indication for the existence and the size of these deviations, independently from the detailed mechanism for it's origin:  Whatever causes the ridge involves degrees of freedom separated in configuration space by $\sim fm$ (the jet cone volume) which however are separated in flow rapidity by the ridge size. (Considerably more than the space time rapidity extent of the jet cone).  It is then likely that such correlations appear in interactions between soft (thermalized) degrees of freedom just as they appear in the interactions between the system and the jet.

It is reasonable to 
assume, as an ansatz, the deviations are Gaussian, and the density in $\eta$ of matter flowing  with rapidity $y$ is
\begin{equation}
\phi(y,\eta) \sim \exp\left[ \frac{-(\eta - y)^2}{2 \sigma_\eta^2} \right]
\end{equation}
where $\sigma_\eta$ is a parameter to be determined.
Putting in this factor instead of the $\delta$-function in Eq. \ref{rhoybgk}, and integrating Eq. \ref{pzbgk} yields, at $\eta=0$ 
\begin{equation}
\label{bgkfluct}
\ave{p_z x_\perp} \sim \frac{1}{2 \sqrt{ 2 \pi}} \left( B   e^{\frac{1}{2} y_L ( -2 - \frac{y_L}{\sigma_\eta^2})} \sigma_\eta \left( 2 - 2 e^{2 y_L} +  \right. \right. \end{equation}
\[\ \left. \left. e^{\frac{(\sigma_\eta^2 - y_L)^2}{\sigma_\eta^2}} \sqrt{2 \pi} \sigma_\eta \left( -\mathrm{erf} \left[ \frac{\sigma_\eta^2 - y_L}{\sqrt{2} \sigma_\eta} \right] + \mathrm{erf} \left[ \frac{\sigma_\eta^2 - y_L}{\sqrt{2} \sigma_\eta} \right] \right) \right)  \right) \]
this somewhat unwieldy expression simplifies,at mid rapidity, to 
\begin{equation}
\label{sigmasimple}
\ave{p_z x_\perp} \propto B e^{\sigma_\eta^2/2} \sigma_\eta^2
\end{equation}
The net-vorticity at mid-rapidity then becomes equal, roughly to the initial
vorticity in the firestreak case multiplied by $e^{\sigma_\eta^2/2} \sigma_\eta^2$ 
and divided by the longitudinal velocity of the nucleons.
\begin{equation}
\label{ratio}
\frac{\left. \ave{P_q^R} \right|_{BGK}}{\left. \ave{P_q^R}
\right|_{firestreak}} = c(s)\frac{m_N e^{\sigma_\eta^2/2} \sigma_\eta^2}{\sqrt{s}}
\end{equation}
\begin{figure*}[t]
\epsfig{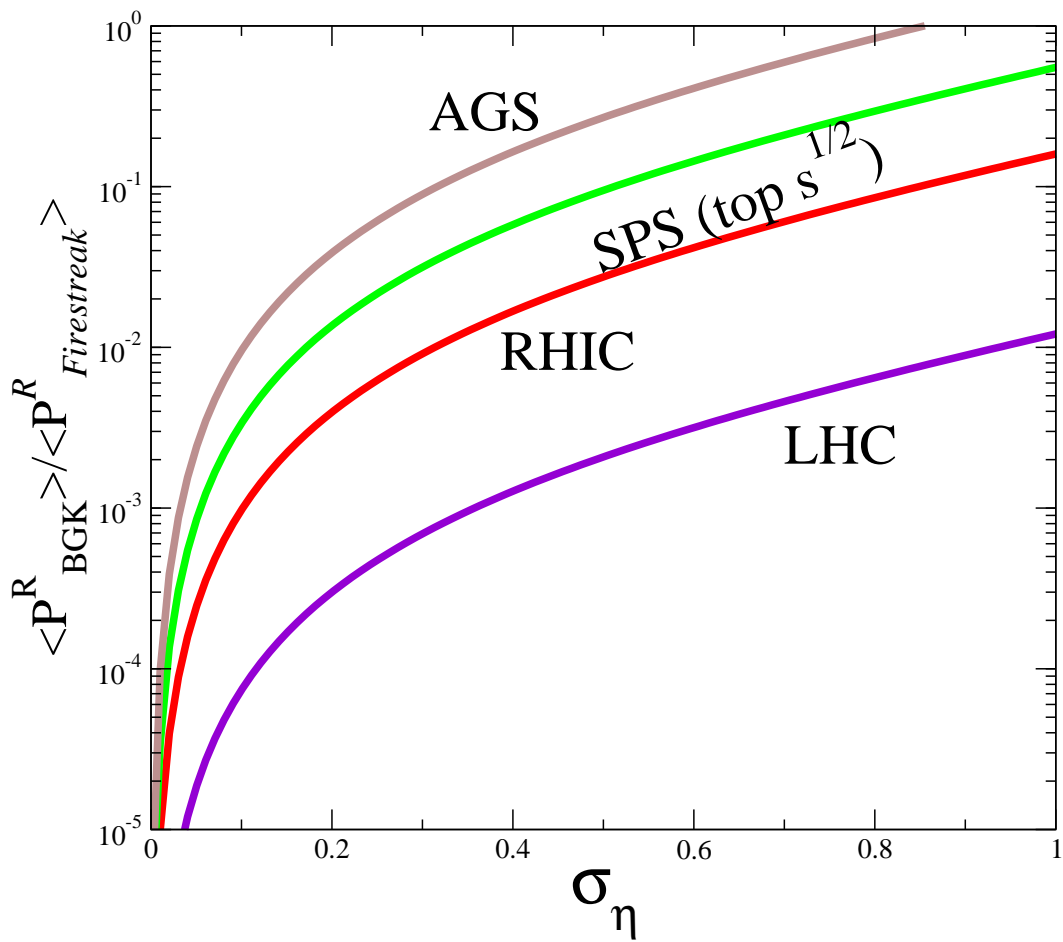}
\caption{\label{bgkfire} (color online)
Ratio of BGK to firestreak predictions as a function of $\sqrt{s}$ and $\sigma_\eta$, the correlation length between spacetime and flow rapidity, calculated using Eqn. \ref{ratio} and \ref{csmilov} }
\end{figure*}
In the limit of $\sigma_\eta \rightarrow 0$ the system has no vorticity. While only at very low energies (where the formula in Eq. \ref{sigmasimple} and the BGK picture are untenable as approximations) the BGK and Firestreak pictures are comparable, vorticity at BGK could still be non-negligible provided $\sigma_\eta \sim 1$.
 
 It should be underlined that $c(s)$ contains very different physics from $\sigma_\eta$:  In \cite{liang}, $c(s)$ is interpreted as the number of partons into which the energy of the initial collision energy is distributed.   $\sigma_\eta$, on the other hand, depends on the imperfection of 
 ``Bjorken'' expansion (correlation between spacetime and flow rapidity).  These two effects, however, go in the same direction, although $c(s) \sim \left( \ln  \sqrt{s} \right)^2 $ is much less efficient at diminishing polarization than a small $\sigma_\eta$.

Combining $c(s)$ of Eq. \ref{csmilov} with Eq. \ref{ratio} we obtain the ratio between BGK and Firestreak expectation, and it's dependence on energy and the parameter $\sigma_\eta$.  The result is shown in Fig. \ref{bgkfire}, assuming $\sigma_\eta \ll y_{L}$.
  The purpose of this figure should be taken as an illustration of the sensitivity of the polarization measure to the longitudinal structure of the initial condition, rather than as a prediction of the polarization in the two models ( as shown in \cite{qm2006}, the small angle approximation used in \cite{liang} is in any case likely to be inappropriate).
As can be seen, the effects of $c(s)$ in the Firestreak picture are comparable to the effects of a non-negligible $\sigma_\eta$ in the BGK picture only at low energies (where the Firestreak picture is thought to work better).  At top RHIC energy, even at $\sigma_\eta$ of one unit, the BGK polarization should be suppressed with respect to the firestreak expectation with about two orders of magnitude.  This grows to several orders of magnitude for LHC energies.

Thus, the measurement of the $\Lambda$ polarization in the reaction plane could be an  valuable tool of the initial 
longitudinal geometry of the system.
At the moment the longitudinal geometry, and in particular how the longitudinal scale varies with energy 
(at what energy, and if, and how do initial conditions go from ``firestreak'' to ``BGK'') , is not well understood \cite{busza}.
This understanding is crucial for both the determination of the equation of state and the viscosity, since 
longitudinal geometry is correlated with the initial energy density, and hence to the total lifetime of the 
system and the time
in which flow observables can form \cite{scaling}.

The measurement of the energy and system size dependence of $\Lambda$ polarization in the reaction plane at 
mid-rapidity could be a significant step in qualitatively assessing the perfection of the fluid, and determining 
at what energy does the system enter fluid-like behavior.

Connecting the experimental measurement of the $\Lambda$ polarization to the initial condition is, however, 
non-trivial, as this observable is sensitive not just to the initial stage but also to the subsequent evolution 
of the system, up to the final freeze-out.

In the next two sections we will qualitatively discuss the effect the later stages will have on the final observable.
We will argue that, while the observable is likely to be modified by the subsequent evolution, a comparison 
of several kinds of polarization could be useful in obtaining information about not only initial conditions, 
but also the mean free path and the freeze-out scenario.

\section{Hydrodynamic evolution, polarization and jets}
In relativistic hydrodynamics, vorticity works somewhat differently than in the non-relativistic limit 
\cite{florkvort,florkvort2}. While in non-relativistic ideal hydrodynamics, the conserved circulation is defined simply as $\vec{\nabla} \times \vec{v}$, relativistically the conserved vorticity is
\begin{equation}
\vec{\Omega} = \vec{\nabla} \times w \gamma \vec{v}\, ,
\end{equation}
where $w$ is the enthalpy per particle.

In the non-relativistic limit, where $ w \simeq m$ and $\gamma=1$ the usual limit is recovered.
In a relativistic fluid with strong pressure and energy density gradients, on the other hand, vortices 
can be created and destroyed even in perfectly smooth initial conditions, such as the BGK case described 
in the previous section.

\begin{figure*}[ht]
 \begin{minipage}[t]{7cm}
 \hspace*{-3.5cm}
 \epsfig{width=8cm,figure=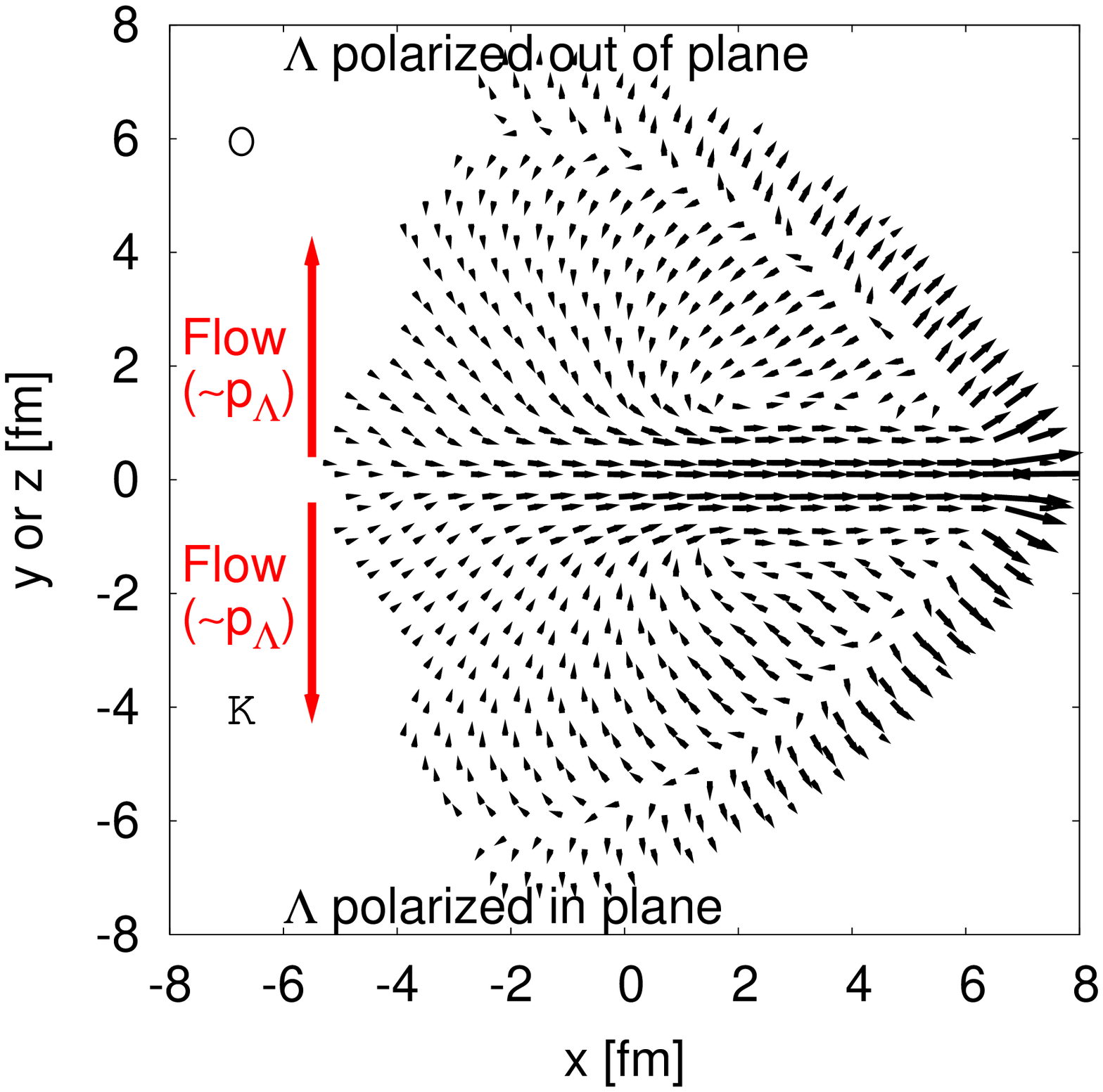}
 \end{minipage}
 \begin{minipage}[t]{7cm}
 \vspace*{-7.7cm}\hspace*{-1.5cm}
 \epsfig{width=8cm,angle=270,figure=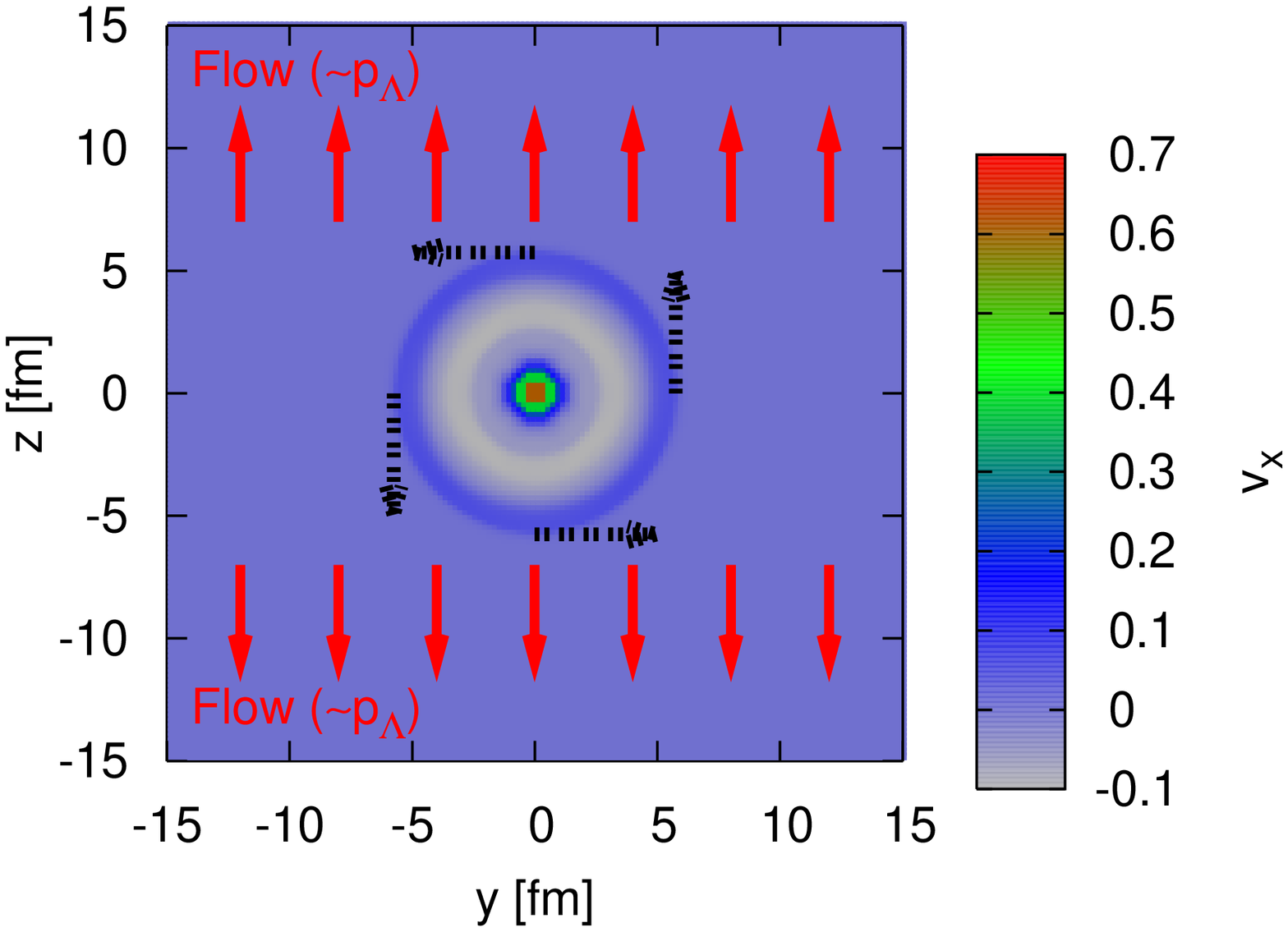}
 \end{minipage}
\caption{\label{smokering} (color online) Vorticity generated by a fast ``jet'' traversing the system in the 
positive x direction.  The arrows in the left panel show the momentum density of fluid elements in the x-y plane, 
while the contour in the right panel shows the x-component of the velocity in the y-z plane. 
The jet has been traveling for $t = 11.52$~fm/c through a static medium \cite{usmach}. The dashed arrows in the right panel indicate the expected direction of polarization of the $\Lambda$ (out of plane for left panel, tangentially in right panel).  If the medium undergoing transverse and longitudinal expansion, the $\Lambda$ position within the smoke-ring is correlated with it's mean momentum.  Thus, measuring $\Lambda$ polarization in the plane defined by its momentum and the jet momentum should yield a positive net result}
\end{figure*}

As vorticity development is a highly non-linear phenomenon, quantitative details require numerical simulations. 
For demonstrative purposes, we include in this paper the vorticity which develops when a momentum source moving at the speed of light traverses a uniform relativistic fluid.  This could be an appropriate description of the thermalized 
jet energy loss, if the jet loses energy fast and locally. The calculation was done using a $(3+1)$D hydrodynamical
code \cite{shasta}. The flow vector in the $x-y-z$ coordinate system where the fluid is at rest (co-moving with the collective flow) is shown in Fig. \ref{smokering}, in parallel (left panel) and perpendicular (right panel) to the jet direction.

The simulation shown in Fig. \ref{smokering} is based on a jet energy loss model that assumes a high 
momentum gradient.   This means either considering the jet as a ``ball of fluid'' moving at ultra-relativistic 
speeds with respect to the medium, or as a source injecting a considerable amount of momentum into the medium.   
The second case is similar to the source found in \cite{shuryak,usmach} to lead to the formation of Mach Cones.     Recent analytical solutions within a strongly coupled N=4 supersymmetric plasma \cite{gubser} show that these vortex-like structures persist in non-equilibrium strongly coupled quantum field theories, suggesting that, if the QGP at RHIC is strongly coupled, they might be found even if the matter surrounding the jet is not quite in local thermal equilibrium.

It is not surprising that a large initial momentum gradient, such as that produced by a jet quickly losing energy, 
can introduce vorticity into the system.
As shown in the included simulation, these vortices are stable enough to last throughout the lifetime of the fluid.
Perhaps, therefore, an interesting polarization measurement to attempt is to trigger on events with jets, and 
measure $\Lambda$ polarization $P_\Lambda^{J}$ in the plane perpendicular to the {\em jet} production plane.   Since vorticity 
in such events exists independently of the global initial conditions, this measurement is sensitive only to the 
mean free path and perhaps final state effects.

Fig. \ref{smokering} also illustrates how such a measurement could be performed: the polarization axis is defined based on the jet (high $p_T$ trigger) direction.  Since vortices above and below the jet move in opposite directions, in a {\em static} medium detecting vorticity via polarization measurements would be impossible.

If, however, the smoke-ring is in a medium undergoing transverse or longitudinal expansion, the flow introduces a correlation between the $\Lambda$ position within the smoke-ring (and hence its polarization) and its average momentum $\ave{p}_\Lambda$.
Measuring the polarization of moderately high momentum but thermal $\Lambda$s ($\sim 700 MeV$) in the plane defined by the $\Lambda$ momentum and the jet direction should therefore yield a non-zero result.

We therefore suggest to measure the polarization $P_i^J$ of jet-associated moderate momentum particles, in the plane defined by the jet direction and the direction of the particle.    The observation of this polarization would be a strong indication of collective behavior, since it would signify jet-induced vorticity.

Unlike production plane vorticity, jet vorticity does not depend on initial condition, but should hold for a wide variety of jet energy loss scenarios, provided the coupling between the system and the jet, and within the system's degrees of freedom, is strong.

It is not at all clear, however, wether in the strongly coupled regime (rather than the perturbative one, on which the calculations of \cite{liang} are based) vorticity will readily transform into quark polarization.   The next section is devoted to this topic.
  
\vspace*{-0.4cm}
\section{Mean free path and polarization \label{sec:lmfppol}}
In a perfect fluid angular momentum should go not into a {\em
  locally} preferred direction, but into vortices where each volume element is locally isotropic in the frame 
  co-moving with the flow.
Such vortices should imprint final observables via longitudinal collective
  flow  ( e.g. odd $v_n$ coefficients away from mid-rapidity), but not via polarization.   In this regime, the equations derived in the first section are no longer tenable since they assume unpolarized incoming particles and a coupling constant small enough for perturbative expansion.  

 Keeping the first of these assumptions would violate detailed balance, while the second assumption is probably incompatible with strong collective behavior.  Thus, if the equilibration between gain and loss terms happens instantaneously (``a perfect fluid''), any created polarization would be instantly destroyed by subsequent re-interactions (In the terminology of eq. \ref{polawang}, $c(s) \rightarrow \infty$, through, since the particles would be ``infinitely strongly'' correlated, this would not imply infinite entropy). The local isotropy of a perfectly thermalized system was used \cite{hoyer} to suggest the disappearance of the production plane polarization observed in elementary collisions could be a signature of deconfinement.
 
  A first order correction
  comes where the size of the radius of curvature within the vortex
  becomes comparable to the mean free path $l_{mfp}$. The anisotropy would then be given by the deformation of a volume element of this size.
  Dimensional
  analysis, together with the insights provided in the first section, yields
\begin{eqnarray}
\ave{P_q^{i}} \sim \tanh\left[ \vec{\zeta}_i \right] \sim \vec{\zeta}_i \label{lmfppol1} \\
\vec{\zeta}_i = \frac{l_{mfp}}{T} \left( \epsilon_{i j k} \frac{d\ave{ \vec{p}_k}}{d \vec{x}_j} \right)  
\label{lmfppol}
\end{eqnarray}
where $\ave{\vec{p_j}}$ is local direction of momentum in the lab frame,$T$ is the temperature, and $i$ is {\em any} direction where a non-zero polarization is expected (production, reaction or jet).  Note that Eq. \ref{lmfppol1} is simply $\ave{\vec{p} \times \vec{x}}$, taken over a homogeneous volume element the size of the mean free path.
Thus, potentially, the amount of residual polarization that survives hydrodynamic evolution 
(wether from initial geometry or from deformation of the system due to jets) is directly connected to the 
system's mean free path.

Measuring the polarization rapidity dependence(in any plane, production, reaction,or jet, where it could be expected to be produced) 
could perhaps ascertain the rapidity domain of the QGP.
If the (s)QGP is formed at central rapidity, while the peripheral
regions are via a hadron gas, one should observe a sharp rise in 
production, reaction and jet plane polarization in the peripheral regions

The problematic aspect of using polarization for such a measurement is that it is sensitive to late-stage 
evolution, including hadronization and the interacting hadron gas phase.

As shown in \cite{hama}, an unpolarized QGP medium at freeze-out
will, through hadronic interactions at last scattering, produce a net
production plane polarization due to the Hadronic
interactions, in a similar way to how unpolarized p-p and p-A collisions result in the net hyperon polarization.  
While local detailed balance will inevitably cancel out such local polarization, the rather large mean free path 
of an interacting hadron gas, and the considerable pre-existing flow ensure that any interacting hadron gas phase 
should be well away from detailed balance, and hence likely to exhibit residual polarization.

It then follows that the absence of production plane polarization would be a strong indication not only of sQGP 
formation, but of a ``sudden'' freeze-out where 
particles are emitted directly from the QGP phase.

The evidence of quark coalescence even at low momentum \cite{coalescence}, together with sudden freeze-out fits 
\cite{florkowski,suddenme}, make this scenario interesting enough to be investigated further using the polarization 
observable in any plane where the vorticity in the hot phase is expected to be non-zero (reaction, production and jet).

If polarization in {\em all} directions is consistently measured to be zero, including events with jets and within high rapidity bins, it would provide strong evidence that the mean free path of the 
system is negligible, and final state hadronic interactions are not important enough to impact flow observables.   
A measurement of production plane but not reaction plane polarization would provide evidence that the initial state 
of the system is BGK-like, and that the interacting hadron gas phase leaves a significant imprint on soft observables.
(The BGK nature of the initial condition can then be further tested by scanning reaction plane polarization in rapidity).

The observation of polarization in the jet plane could provide a further estimate of the mean free path, and show that 
the jet degrees of freedom are thermalized and part of the collective medium.
A sudden jump in any of these polarizations at a critical rapidity could signal a sharp increase in the mean free 
path, consistent with the picture of a mid-rapidity QGP and a longitudinal hadronic fragmentation region.
Analogously, a drop of polarization while scanning in energy and system size could signal the critical parameters required for a transition from a 
very viscous hadron gas to a strongly interacting quark-gluon liquid.
 
In conclusion, we have made a few considerations regarding the physics relevant for hyperon polarization measurements 
in heavy ion collisions.
We have argued that polarization physics is directly connected to several of the more contentious and not understood 
aspects of the system produced in heavy ion collisions, such as initial longitudinal geometry and microscopic 
transport properties.   We have, however, shown that the polarization observable can be significantly altered by all 
of the stages of the system's evolution, thereby rendering a quantitative description of it problematic.
We have argued that measuring polarization in several directions (reaction plane, production plane and jet plane), 
and it's excitation function and rapidity 
domain, could shed some light on the described ambiguities.

\section*{Acknowledgments}
The computational resources have been provided by the Center for 
Scientific Computing, CSC at Frankfurt University.
G. T. thanks the Alexander von Humboldt Foundation for financial 
support. This work has also been
financially supported by GSI and BMBF.
We thank X.N. Wang, F. Becattini,  Ilya Selyuzhenkov, D. Rischke and H. St\"ocker for 
helpful 
discussions.

\end{document}